\documentclass[aps,preprint]{revtex4}

\usepackage{amsbsy}
\usepackage{amssymb}
\usepackage{amsmath}
\usepackage{graphicx}

\def\x{{\mathrm{x}}}
\def\y{{\mathrm{y}}}

\def\x{{\mathrm{x}}}
\def\y{{\mathrm{y}}}

\def\n{{\rm n}}
\def\p{{\rm p}}
\def\e{{\rm e}}

\def\e{{\rm e}}
\def\s{{\rm s}}

\def\N{{\mathrm N}}

\def\be{\begin{equation}}
\def\ee{\end{equation}}
\def\beq{\begin{equation}}
\def\eeq{\end{equation}}
\def\bea{\begin{eqnarray}}
\def\eea{\end{eqnarray}}

\def\bear{\begin{eqnarray}}
\def\eear{\end{eqnarray}}

\newcommand{\Bcal}{\mathcal{B}}
\newcommand{\Acal}{\mathcal{A}}

\newcommand{\dettK}{\mathrm{det}\,\tilde K}

\begin{document}

\title{A minimal model for finite temperature superfluid dynamics}

\author{N. Andersson$^1$,  C. Kr\"uger$^1$, G.L. Comer$^2$ \& L. Samuelsson$^3$}

\affiliation{
$^1$ School of Mathematics, University of Southampton,
Southampton SO17 1BJ, United Kingdom\\
$^2$ Department of Physics and the Center for Fluids at All Scales, St Louis University, USA\\
$^3$ Nordita, Roslagstullsbacken 23, SE-106 91 Stockholm, Sweden}

\begin{abstract}
Building on a recently improved understanding of the problem of heat flow in general relativity, we develop a hydrodynamical model for coupled finite temperature superfluids. The formalism is designed with the dynamics of the outer core of a mature neutron star (where superfluid neutrons are coupled to a conglomerate of protons and electrons) in mind, but the main ingredients are relevant for a range of analogous problems. The entrainment between material fluid components (the condensates) and the entropy (the thermal excitations) plays a central role in the development. We compare and contrast the new model to previous results in the literature, and provide estimates for the relevant entrainment coefficients that should prove useful in future applications. Finally, we consider the sound-wave propagation in the system in two simple limits, demonstrating the presence of second sound if the temperature is sub-critical, but absence of this phenomenon above the critical temperature for superfluidity.  
\end{abstract}

\maketitle

\section{Introduction}

A realistic model of a mature neutron star must account for various extreme aspects of physics. In addition to the strong gravity, we need to consider the state of matter at supranuclear densities.  We also need to ask how the composition of matter changes as the density increases. Many different options, involving varying levels of exotica, have been suggested. For example, at the present time we do not know if the deep neutron star core contains deconfined quarks, perhaps in the form of a colour superconductor \cite{CSC}. Meanwhile, the composition of the outer core is better understood \cite{LP}. Just above the crust-core transition density (at around 60\% of the nuclear saturation density), the matter is predominantly made up of neutrons with a small fraction of protons and electrons (to make the conglomerate charge neutral). As the density increases muons appear, eventually followed by (this is where uncertainties creep in) hyperons and/or quark deconfinement. As a result, a neutron star is a bit like a layer cake where the composition of the different layers affects the transport coefficients which in turn determine the effect on cooling and the viscous damping of oscillations. Key to  modelling this physics in a realistic fashion is the fact that several regions in the star are likely to exhibit superfluidity/superconductivity.

\begin{figure}
\includegraphics[height=10cm]{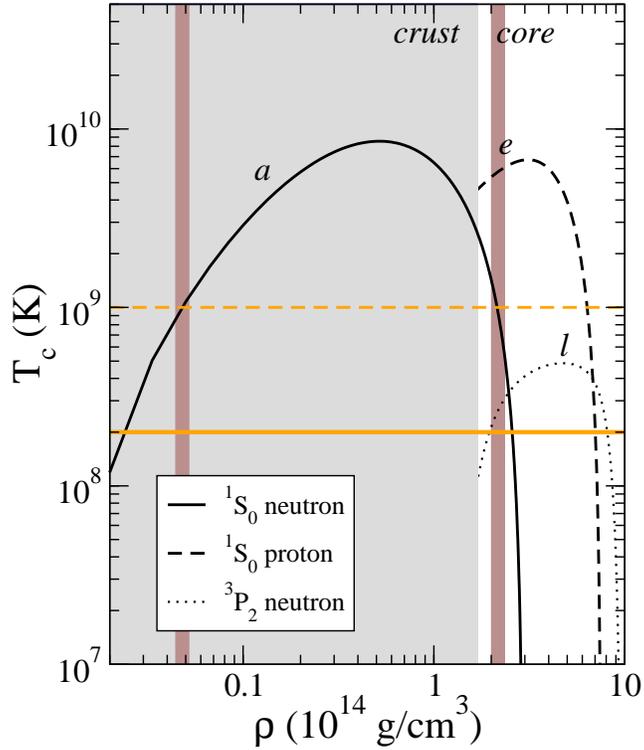}
\caption{A schematic illustration of the critical temperatures $T_c$ for superfluidity of neutrons (singlet: solid line, triplet: dotted line) and protons (singlet: dashed line) for representative pairing gaps, cf. \cite{nuclpa}. Each critical temperature is shown as function of the total (energy) density $\rho$. The neutron star crust region at lower densities is indicated in grey. Examples of the critical densities at which the superfluid transition takes place for a given temperature are indicated as horizontal (orange) lines. The regions where thermal excitations are expected to have a considerable effect on the dynamics of the neutron component are shown as vertical (darker) bands for a sample temperature of $10^9$~K.}
\label{schemes}
\end{figure}

If we focus our attention on neutron star oscillations, then the added degree of freedom associated with superfluidity leads to the presence of extra families of oscillation modes \cite{Epstein,Lindb,Lin,ACmnras} (analogous to the second sound in laboratory superfluids). This aspect of the problem has already been considered at some level of detail. However, most previous models ignore finite temperature effects. This may seem natural given that a typical 
neutron star is cold (in the sense that the temperature is orders of magnitude below the neutron Fermi temperature). While this statement is certainly true, we also need to acknowledge that the various superfluid pairing gaps are density dependent \cite{nuclpa}. The upshot of this is that there should 
\underline{always} be regions in a mature neutron star that are near the superfluid transition, see Figure~\ref{schemes}. In such a region finite temperature effects will dominate. These transition regions may be (highly) localised, but it would nevertheless be wise to consider them in detail. They may, in fact, have an unexpectedly large influence on the dynamics of the system. Consider for example the analogy with the viscous boundary layer at the crust-core interface and the suggested effect on the gravitational-wave driven r-mode instability \cite{Ek1,Ek2,Ek3,Ek4}. In this case it is the finite extent of the boundary layer that leads to it having a large effect on the dynamics. Thus, without considering the problem in more detail we cannot know if the superfluid transition region is dynamically important or not. 

In order to address this problem, we need a model that accounts for finite temperature effects in neutron star superfluids. Such models have been discussed, in particular by Gusakov and collaborators \cite{gus1,gus2,gus3}, but the problem needs further attention. Our aim is to develop a phenomenological model that (i) enables us to consider finite temperature superfluid dynamics, and at the same time is (ii)
flexible enough to allow for future extensions involving (for example) the star's magnetic field and/or the elastic crust.
For obvious reasons the discussion will focus on the matter components, especially the extra degrees of freedom that enter in a superfluid system.
Still, one should keep in mind that the complete model involves the ``live'' spacetime of Einstein's theory. We will not discuss the associated
(gravitational-wave) aspects here, but it is important to remember that the matter dynamics that we consider is coupled to the spacetime geometry.
In other words, the spacetime metric $g_{ab}$ is not taken to be fixed in the following. In any realistic application, one also has to solve (at least a subset of) the Einstein field equations.

As we intend to develop a covariant model, we distinguish between co- and contravariant quantities using the space-time metric to relate the two. Spacetime indices are given as italic letters from the beginning of the alphabet, $a, b, c, ...$. In cases where the focus is on spatial components the corresponding indices are $i, j, k, ...$. We assume that the spacetime signature is +2 and the Einstein summation convention applies. The spacetime indices should be distinguished from the constituent indices that are used to identify the different fluid components in the problem. These are given as roman, generically x, y, z,... and  specifically (in the present context) n, p, e and s. Repetition of these component labels in a mathematical expression does not imply summation.

\section{The variational approach}

In contrast to the deep core, the conditions in the outer core of a neutron star are relatively well understood. In this region a (triplet pairing) neutron superfluid is thought to coexist with a (singlet pairing) type II proton superconductor \cite{supercon} and a sea of relativistic electrons (we will ignore muons in the following, even though they are quite easy to account for). In a hot system we also need to account for the entropy. In general, all four components can flow relative to one another. A general model would therefore consider four distinct fluid components; 
neutrons (labelled n in the following), protons (p), electrons (e) and the entropy (s). In order to focus on the finite temperature effects, we will ignore two of the relative degrees of freedom. The first represents the charge current due to a relative flow between protons and electrons (as required to maintain the global magnetic field \cite{namhd}), and the second corresponds to the heat flux \cite{cesar1,cesar2}. In principle, both of these aspects can be included in the model (at the cost of making the involved relations less manageable \cite{namhd}) but  the added complexity detracts from the key points of the analysis. Hence, we first of all assume that the protons and electrons are locked electromagnetically. In the final model we also add the assumption that the entropy is carried along with the only ``normal'' component in the system, the electrons.  This leaves us with two fluid degrees of freedom. This should be the minimal model for the system of interest.

We take as our starting point the convective variational principle for multi-fluid systems devised by Carter \cite{carter} (see \cite{livrev} for a review).
In this model, the fluid dynamics is obtained from a matter Lagrangian $\Lambda$, which is taken to be a function of the various scalars that can be constructed from the individual fluxes $n_\x^a$, where the constituent index $\x=\n,\p,\e$ or $\s$. Basically, this means that $\Lambda$ is a function of
\beq
n_\x^2 = - n^\x_a n_\x^a \ , \qquad \mbox{ and} \qquad 
n_{\x\y}^2 = - n^\x_a n_\y^a \ .
\eeq
As will become apparent later, the second of these definitions represents effects due to the relative flow between the different components. While this flow is generally expected to be
small in magnitude, its contribution is nevertheless significant since it encodes the entrainment effect \cite{ent1,ent2,ent3}.  This effect relies on quadratic (and higher order) terms in the Lagrangian. Hence, it is important that we do not make immediate use of the low relative velocities. This assumption should not be incorporated until the relevant momenta have been defined. 

Given the general form for $\Lambda$, the variational procedure determines the
momenta that are canonically conjugate to the  fluxes \cite{livrev}. We have
\beq
  \mu_a^\x = \frac{\partial \Lambda}{\partial n^a_\x} = \Bcal^\x n_a^\x + \sum_{\y\neq\x} \Acal^{\y\x} n_a^\y \ , \label{mu}
  \ee
where
\begin{align}
  \Bcal^\x     &= -2\frac{\partial \Lambda}{\partial n_\x^2} \ , \\
  \Acal^{\y\x} &= -\frac{\partial \Lambda}{\partial n_{\y\x}^2} \ .
\end{align}
The latter encodes the entrainment between the different components. We will account for two distinct entrainment mechanisms. First of all, we 
 know that the entrainment between neutrons and 
protons arises due to the strong interaction. Hence, we will have $\Acal^{\p\n}\neq0$. We also know that the entropy entrainment plays a central role in ensuring causality of the heat flux \cite{cesar1,cesar2}. Hence, we assume that $\Acal^{\s\x}\neq 0 $ (with $\x=\n,\p$), as well. We will not account for entrainment between electrons and baryons. 

It is useful to distinguish the entropy flux $s^a = n_\s^a$ and the associated thermal momentum  $\theta_a = \mu^\s_a$ \cite{cesar1,cesar2} from the other components. This makes the description slightly less ``symmetric'' but it makes it easier to make contact with thermodynamics.   We then have 
\beq
\theta_a = \frac{\partial \Lambda}{\partial s^a}  = \Bcal^\s s_a + \sum_{\y\neq\s} \Acal^{\y\s} n_a^\y \ . 
\eeq

Finally,  a variation with respect to the spacetime metric tells us that  
the stress-energy tensor for the system takes the form
\beq
  T^a{}_{b} = \Psi\delta^a{}_{b}
     +  \sum_\x n_\x^a\mu_b^\x \ ,
\eeq
where the generalized pressure is defined as
\beq
  \Psi = \Lambda - \sum_\x n_\x^a\mu_a^\x \ .
\eeq

The stress-energy tensor acts as source in Einstein's equations in the usual way. In the case where the fluxes are individually conserved, i.e.,
when the dynamical timescale is much faster than that associated with the relevant reactions, we also have
\beq
\nabla_a n_\x^a = 0  \ , \qquad 
\nabla_a s^a = 0 \ .
\eeq
In this (the fully variational) case, each component satisfies an equation of motion of  form
\be
  2n^a_\x\nabla_{[a}\mu^\x_{b]} = 0 \ ,   
\label{eul}  \ee
  and
\be
  2s^a\nabla_{[a}\theta_{b]} = 0 \ . 
  \label{euler}
\ee
These Euler equations combine to  ensure the conservation
of energy and momentum, $\nabla^aT_{ab} = 0$. 

As in the case of relativistic heat flux \cite{cesar1,cesar2}, it is useful to focus on a particular fluid frame. Basically, we need to decide which observer is measuring the temperature etcetera.  In the present context, 
it seems natural to work in the entropy frame (associated with the flow of the ``normal'' fluid). Hence, we introduce an observer with four velocity 
$u^a$ such that $s^a=su^a$. 
In order to quantify the relative flow associated with the superfluid components, we define the relative velocities $v_\x^a$ as
\beq
  n_\x^a = \gamma_\x n_\x(u^a + v_\x^a) \ ,
  \qquad u^av^\x_a = 0 \ ,
  \qquad \gamma_\x = (1-v_\x^2)^{-1/2} \ ,
  \qquad v_\x^2 = v_\x^av^\x_a \ .
\eeq
In many cases of interest $v_\x^2$ will be small, and  it makes sense to work with an expansion with these ``drift velocities'' as small parameters. Rather than working with the individual velocities, it is often convenient to introduce the relative velocities $w_{\x\y}^a = v_\x^a - v_\y^a$. Then the entrainment parameter depends on
\be
n_{\x\y}^2 \approx n_\x n_\y \left( 1 + {w_{\x\y}^2 \over 2} \right) \ , 
\label{relvel}\ee
and we see that the equation of state, $\Lambda$, must contain information at order $w_{\x\y}^2$ in order for the momenta, $\mu^\x_a$, to be correct at linear order. Problems that require higher order information must be considered on a case-by-case basis. For example, in the Hartle-Thorne slow-rotation approximation for a two-fluid system \cite{nilsgreg} (accurate to order $w_{\x\y}^2$) one can show that contributions originating from order $w_{\x\y}^4$ terms cancel exactly \cite{greg}

In the general case, the number densities measured in the entropy frame (such quantities will be denoted by $*$ in the following) are
\be
n_\x^* = - u_a n_\x^a = \gamma_\x n_\x \ .
\ee
We then have the momenta (associated with the three different fluxes)
\be
  \mu_a^\n = \Bcal^\n n_a^\n +  \Acal^{\p\n} n_a^\p +  \Acal^{\s\n} s_a = \left( \Bcal^\n n_\n^* + \Acal^{\p\n} n_\p^* +  s  \Acal^{\s\n} \right) u_a +  \Bcal^\n n_\n^* v^\n_a + \Acal^{\p\n} n_\p^* v^\p_a \ ,
\label{fifteen}\ee
\be
  \mu_a^\p = \Bcal^\p n_a^\p +  \Acal^{\n\p} n_a^\n +  \Acal^{\s\p} s_a = \left( \Bcal^\p n_\p^* + \Acal^{\n\p} n_\n^* +  s  \Acal^{\s\p} \right) u_a +  \Bcal^\p n_\p^* v^\p_a + \Acal^{\n\p} n_\n^* v^\n_a \ ,
\ee
and
\be
\theta_a = \Bcal^\s s_a +  \Acal^{\n\s} n_a^\n +  \Acal^{\p\s} n_a^\p =  \left( \Bcal^\s s + \Acal^{\n\s} n_\n^* +   \Acal^{\p\s} n_\p^* \right) u_a +
 \Acal^{\n\s} n_\n^* v^\n_a  +   \Acal^{\p\s} n_\p^* v^\p_a \ .
\ee
Note that we have not yet locked the electrons (which carry  entropy and therefore move along with the entropy component) to the protons (we still have $v_\p^a\neq 0$).
Given these expressions we see that the ``chemical potentials'' (in the entropy frame) are
\be
\mu_\n = -u^a \mu^\n_a = \Bcal^\n n_\n^* + \Acal^{\p\n} n_\p^* +  s  \Acal^{\s\n} \ , 
\label{chempot}\ee
\be
\mu_\p = -u^a \mu^\p_a = \Bcal^\p n_\p^* + \Acal^{\n\p} n_\n^* +  s  \Acal^{\s\p} \ , 
\ee
and
\be
\mu_\e = -u^a \mu^\e_a = \Bcal^\e n_e \ , 
\ee
while the temperature $\theta$ follows from
\be
\theta = - u^a \theta_a = \Bcal^\s s + \Acal^{\n\s} n_\n^* +   \Acal^{\p\s} n_\p^* \ . 
\ee

We can use these expressions to rewrite the stress-energy tensor and the pressure. This is not necessary but it clarifies the physical meaning of the variational coefficients. We need
\be
n_\n^a \mu^\n_a = - n_\n^* \mu_\n  + n_\n^* \left( \Bcal^\n n_\n^*  v_\n^2 + \Acal^{\p\n} n_\p^* v_\n^a v^\p_a\right) \ , 
\ee
\be
n_\p^a \mu^\p_a = - n_\p^* \mu_\p  + n_\p^* \left( \Bcal^\p n_\p^*  v_\p^2 + \Acal^{\n\p} n_\n^* v_\p^a v^\n_a\right) \ , 
\ee
\be
n_\e^a \mu^\e_a = - n_\e \mu_\e \ , 
\ee
and
\be
s^a \theta_ a = - s \theta \ .
\ee

Given these results we have
\be
\Psi = \Lambda +  n_\n^* \mu_\n  +  n_\p^* \mu_\p +  n_e \mu_e + s \theta-  \left[ \Bcal^\n \left( n_\n^*  v_\n\right)^2 + 2 \Acal^{\p\n} n_\n^* n_\p^* v_\n^a v^\p_a +  \Bcal^\p \left( n_\p^*  v_\p\right) ^2 \right] \ . 
\ee
This is the (extended) Gibbs relation for the system, accounting for the presence of three distinct flows.

Let us now explicitly lock the protons to the electrons, i.e. assume that $v_\p^a=0$. Then we are left with 
\be
\Psi = \Lambda +  n_\n^* \mu_\n  +  n_\p \mu_\p +  n_\e \mu_\e + s \theta-  \Bcal^\n \left(n_\n^*  v_\n\right)^2 \ . 
 \ee
 Finally, in most situations of practical interest the deviation from co-motion will be small. When that is the case we can neglect terms of
 order $v_\n^2$, leaving us with
\be
\Psi = \Lambda +  n_\n \mu_\n  +  n_\p\mu_\p +  n_\e \mu_\e + s \theta \ . 
\label{Gibbs} \ee
 In the following we will assume that these various approximations are valid for the neutron star problem. Should we be interested in a more general context, it is 
 straightforward to reinstate the redshift factors, etcetera.
 
Given the above relations, we can work out the corresponding form for the stress-energy tensor. To do this, we need  
\begin{multline}
\sum_\x n_\x^a \mu^\x_b \approx \left( \Psi-\Lambda\right)u^a u_b + n_\n\mu_\n v_\n^au_b
+ n_\n \left( \Bcal^\n n_\n + \Acal^{\n\p} n_\p +  \Acal^{\n\s} s\right) u^a v^\n_b \\
= \left( \Psi-\Lambda\right)u^a u_b + n_\n\mu_\n \left(v_\n^au_b + v^\n_b u^a \right)  \ . 
\end{multline}
The final expression for the stress-energy tensor is 
\be
T^a_{\ b} = \Psi \delta^a_{\ b} + \left( \Psi-\Lambda\right)u^a u_b + n_\n\mu_\n \left(v_\n^au_b + v^\n_b u^a \right) = 
-\Lambda u^a u_b + \Psi \perp^a_{\ b} + n_\n\mu_\n \left(v_\n^au_b + v^\n_b u^a \right) \ , 
 \ee
 where we have used the standard projection
 \be
 \perp^{ab}= g^{ab} + u^a u^b \ .
 \ee
 It is worth noting that the relative flux enters $T^a_{\ b}$ in the same way as the heat flux would \cite{cesar1,cesar2}. This is, of course, as expected.

Noting that the energy density measured by an observer moving along with the entropy-proton-electron conglomerate is
\be
\rho = u_a u_b T^{ab} = -\Lambda \ , 
\ee 
we recognize the Gibbs relation \eqref{Gibbs} as the \underline{definition} of the pressure $\Psi=P$, with $\theta=T$ the temperature. In other words,  in the linear case our final expressions are
\be
T^a_{\ b} = \rho u^a u_b + P \perp^a_{\ b} + n_\n\mu_\n \left(v_\n^au_b + v^\n_b u^a \right) \ , 
\label{seten} \ee
 and
 \be
 P+\rho = n_n \mu_\n + n_\p \mu_\p + n_\e\mu_\e + sT \ . 
 \ee
 
The simplified problem has \underline{two} dynamical degrees of freedom (the remaining fluxes). To determine these we need two momentum equations, e.g. \eqref{eul} and \eqref{euler}. However, if we want to compare to the standard single-fluid problem, then it makes sense to first of all consider
\be
\nabla_b T^{ab} = 0 \ , 
\ee
representing the conservation of total energy and momentum.
In addition, we will use the Euler equation for the neutrons
\be
 2n^a_\n\nabla_{[a}\mu^\n_{b]} = 0 \ , 
\label{nmom}\ee
where, cf., Eq.~\eqref{fifteen}, 
\be
\mu^\n_a =\mu_\n u_a +  \Bcal^\n n_\n v^\n_a  \ .
\label{momn}\ee

So far, the neutron superfluidity is only manifest from the fact that the corresponding component is allowed to flow relative to the other fluids in the system, i.e. there is no viscous coupling. This is the generic model one arrives at in situations where quantized vortices are present in the system. The macroscopic equations are obtained by averaging over the vortices. In order to make contact with the standard analysis of laboratory systems, it is useful to consider the fact that a ``pure'' superfluid should be  irrotational. In this case the flow is potential,  such that we have the momentum
\be
\mu^\n_a = m_\n \nabla_a \phi_\n \ , 
\ee
for some scalar potential $\phi_\n$ (proportional to the order parameter of the condensate). In this case \eqref{nmom} 
is automatically satisfied. Later, we need to make contact with the ``orthodox'' model for superfluidity, centered on the notion of a superfluid (spatial) ``velocity'' $V_\n^i$. This quantity represents the scaled momentum of the superfluid component \cite{livrev}, and (in the entropy frame) we have
\be
V_\n^i \equiv  {1\over m_\n} \nabla^i \phi_\n \ .
\ee
From \eqref{momn} we then see that we must have
\be
\mu_\n = -u^a \mu^\n_a = -m_\n u^a \nabla_a \phi_\n \ , 
\ee
and
\be
V_\n^a = {1 \over m_\n }\perp^a_{\ b} \mu_\n^b =  {\Bcal^\n n_\n \over m_\n} v_\n^a  \ .
\ee
This shows that the ``velocity'' $V_\n^a$ is parallel, but not identical, to the relative  velocity $v_\n^a$. As we will see later, the difference is associated with the entrainment effect (arising when the neutron effective mass differs from the bare mass).

\section{Newtonian correspondence and effective masses}

The model we have developed is complete, at least formally. However, in neutron star modelling there tends to be a void between theory formulations and applications. The main reason for this is that the true supranuclear equation of state remains elusive. To date, most discussions have focussed either on bulk properties (as required to construct stationary self-gravitating models, leading to mass and radius \cite{LP}) or the transport coefficients associated with cooling \cite{cool} (reaction rates etcetera). Comparatively little attention has been given to the coefficients required to model superfluid dynamics, like the entrainment or additional dissipation coefficients \cite{monster,formal}. The notable exceptions are the pioneering work  of Alpar et al \cite{alpar} and Borumand et al  \cite{boromir} at the Newtonian level and Comer and Joynt \cite{hobbits} in the relativistic case. The most detailed recent work is that of Chamel and collaborators \cite{chamel1,chamel2,chamel3} for the neutron star crust and  Gusakov et al for hyperon cores \cite{Gusk1,Gusk2}. It is also worth noting recent discussions of the additional bulk viscosities that arise in superfluid systems \cite{gusbulk,guskan,cristina} . Another important point concerns the fact that many quantum equation of state calculations are carried out in a non-relativistic framework. This is natural since the nuclear physics scale is always small enough that one can work in a local inertial frame \cite{glenn}. However, it means that  we need to be able to implement non-relativistic equation of state parameters in our fully relativistic framework for superfluid dynamics if we want to make progress  (given the available information).

To discuss this issue, it is useful to  first focus on a zero temperature model. Then we are dealing only with two components and many of the issues become conceptually clearer. Moreover, as a first stab at the finite temperature problem it makes sense to build on the cold model, simply assuming that the thermal effects (say) the strong interaction are small. We develop this strategy through this section and then return to the finite temperature problem. 

Let us consider the low-relative velocity version of the variational model. Letting the two components be the neutrons and the ``protons" (n and p, respectively, with the electrons locked to the protons) and writing the variation of $\Lambda$ in such a way that the relative velocity $w^2=w_{\n\p}^2$ is explicit, we have [making use of \eqref{relvel}]
\begin{multline}
d\Lambda = - \left[ n_\n \Bcal^\n + n_\p \left( 1+{w^2 \over 2 c^2} \right) \Acal^{\n\p} \right] dn_\n  \\
- \left[ n_\p \Bcal^\p + n_\n \left( 1+{w^2 \over 2 c^2} \right) \Acal^{\n\p} \right] dn_\p 
-{n_\n n_\p \over 2 c^2} \Acal^{\n\p} dw^2 \ . 
\label{dL}\end{multline}
As we want to consider the Newtonian limit of this expression, we have reinstated the speed of light (which is taken to be unity in the rest of the discussion).  
In considering the limit $c\to \infty$ we use units such that  the time coordinate is $x^0 = ct$ and the spatial components of the four velocity $u_\x^i\to v_\x^i/c$. With these conventions it is still the case that $n_\x^2 = - n_\x^a n^\x_a$ has the dimension of a number density.
The analysis is also simplified by the fact that we 
are already considering low relative 
velocities (the detailed calculation has been carried out in, for example, \cite{livrev} and we refer the interested reader to that discussion for more details). The key point is that the total energy of the relativistic system includes the rest-mass contributions, but these are not present at the Newtonian level. In effect, we have
\be
\Lambda = - \sum_\x m_\x n_\x c^2 - \mathcal{L} \ , 
\ee
where $\mathcal{L}$ represents the relevant Newtonian Lagrangian \cite{monster} (and the relative sign is convention).
The Newtonian momenta follow from \cite{monster}
\be
p^\x_i = \left( {\partial \mathcal{L} \over \partial n_\x^i }\right)_{n_\x} = K^{\x\x} n^\x_i + K^{\x\y} n^\y_i \ , 
\ee
where we have expressed $\mathcal{L}$ in terms of the mobility matrix introduced by Chamel \cite{chamass}. That is, we define $K^{\x\y}$ such that
\be
\mathcal{L} = {1\over 2} \sum_{\x\y} g_{ij} K^{\x\y}n_\x^i n_\y^j+\mathcal{U}(n_\x)  \qquad \x=\n,\p \ , 
\ee
where the fluxes $n_\x^i=n_\x v_\x^i$ are (obviously) spatial (relative to $u^a$) and $\mathcal{U}$ depends on the internal energy energy. 

Given these definitions we return to \eqref{dL}, and find that 
\be
\left( {\partial \Lambda \over \partial w^2 }\right)_{n_\x} = - {n_\n n_\p \over 2c^2} \Acal^{\n\p} \to -  \left( {\partial \mathcal{L} \over \partial w^2 }\right)_{n_\x} = 
 - {n_\n n_\p \over 2} K^{\n\p} \ . 
\ee
Hence, it makes sense to identify
\be
\Acal^{\n\p} = c^2 K^{\n\p} \ . 
\ee
Similarly, we get 
\be
\left( {\partial \Lambda \over \partial n_\n }\right)_{n_\p,w^2} = - n_\n \Bcal^\n - n_\p \left( 1 + {w^2 \over 2 c^2} \right) \Acal^{\n\p}  = -m _\n c^2 - \mu_\n^\N \ , 
\ee
where we have defined the Newtonian chemical potential as
\be
\mu_\n^\mathrm{N} = \left( {\partial \mathcal{U} \over \partial n_\n} \right)_{\n_\p, w^2} \ . 
\ee 
At the level of linear relative velocities, we identify
\be
n_\n \Bcal^\n = \left( m_\n -n_\p K^{\n\p}  \right) c^2 + \mu_\n^\mathrm{N} \ . 
\ee

To leading order, the relativistic momentum leads to
\be
\mu_\n = n_\n \Bcal^\n + n_\p \Acal^{\n\p} = m_\n c^2 + \mu_\n^\mathrm{N} \ , 
\label{chemrel}\ee
and 
\be
\mu_\n^i = \left( m_\n -n_\p K^{\n\p}  \right) c v_\n^i +  n_\p K^{\n\p} c v_\p^i  = c p_\n^i \ , 
\ee
(the factor of $c$ is required to balance the fact that $x^0=ct$),
from which we see that 
\be
n_\n K^{\n\n} =  m_\n -n_\p K^{\n\p} \ . 
\ee

The entrainment has an intuitive interpretation in terms of an effective (dynamical) mass. This quantity follows by considering the neutron momentum in the proton frame (where $v_\p^i=0$), and vice versa \cite{ent3}. That is, we have
\be
m_\n^*=  m_\n -n_\p K^{\n\p}
\quad \longrightarrow \qquad  n_\p K^{\n\p} = m_\n - m_\n^* \ , 
\ee
which means that 
\be
n_\n K^{\n\n} = m_\n^* \ . 
\ee
A similar analysis for the protons leads to 
\be
n_\p K^{\p\p} = m_\p^* \ , 
\ee
and
\be
 n_\n K^{\n\p} = m_\p - m_\p^* \ . 
\ee
From this  we deduce that that the two effective masses are related:
\be
m_\n - m_\n^* = {n_\p \over n_\n} \left( m_\p - m_\p^* \right) \ . 
\ee

Before we proceed, it is worth noting that we can (obviously) define an analogous effective mass $\bar{m}_\n^*$ in the relativistic model. 
In this case, the argument leads to [using \eqref{chemrel}]
\be
\bar{m}_\n^* = n_\n \Bcal^\n = \mu_\n + (m_\n^* - m_\n)
\ee
This relation shows that, in the absence of interactions (read, entrainment), the effective relativistic mass is equal to the chemical potential. This is as expected.

Summarising the discussion, we have shown how a relativistic model can be ``inferred'' from the Newtonian limit in the sense that we can obtain $\Bcal^\x$ and $\Acal^{\x\y}$ from the rest masses $m_\x$, the effective masses $m_\x^*$ and the chemical potentials $\mu_\x^\mathrm{N}$ (which obviously follow from the energy functional $\mathcal U$). The procedure is not ``unique'', but we have to resort to it in lieu of relativistic equation of state models. In particular, the analysis  provides the translation between the variational model and the non-relativistic microphysics, e.g. the information encoded in the entrainment. The relevant information is expressed in terms of a single parameter, which can be taken to be the effective mass $m_\p^*$.

We now want to build on this strategy, extending it to account for finite temperature effects. To do this, we need to make contact with the ``orthodox'' framework for superfluidity, because the problem has so far only been considered in that context  \cite{gusha1,gus1}. 
In other words, we want to compare our results to ones based on splitting each condensate into a ``normal'' part and a ``superfluid'' part (each expressed in terms of a mass density). This language provides a convenient description of phenomena observed in laboratory measurements. The variational model contains exactly the same information \cite{helium}, although the system is decomposed in a different fashion. The preference of one model over the other is essentially a matter of personal choice. After all, one cannot truly decouple the involved degrees of freedom (e.g. the entropy and the nucleons, or the superfluid fraction and the condensate). They are inextricably linked. 

As already mentioned, the orthodox model focuses on the superfluid velocities (the rescaled momenta). In the Newtonian case,  we have $V^\x_i = p^\x_i/m_\x$ which leads to a total momentum flux
\be
j_\x^i = \rho_\x v_\x^i = \rho_{\x\x} V_\x^i + \rho_{\x\y} V_\y^i\ , \qquad \y \neq \x \ , 
\ee
where the entrainment is now expressed in the elements of the mass density matrix $\rho_{\x\y}$ \cite{ent1,ent2}.
Comparing to the variational model, we see that (after a bit of algebra and setting $m_\n = m_\p=m$ for simplicity)
\be
K^{\n\n} = {m^2 \rho_{\p\p} \over \mathrm{det}\ \rho } \ , \qquad K^{\p\p} = {m^2 \rho_{\n\n} \over \mathrm{det}\ \rho }  \ , \qquad \mathrm{and} \qquad  K^{\n\p} = -{m^2 \rho_{\n\p} \over \mathrm{det}\ \rho }  \ , 
\label{Krel}\ee
where
\be
 \mathrm{det}\ \rho = \rho_{\p\p} \rho_{\n\n}- \rho_{\n\p}^2  \ . 
\ee
For future reference, it is  worth noting that we can relate the mass density matrix to the effective masses as well. We then have
\be
\rho_{\n\n} = \rho_\n m_\p^* \left[ m_\p^* - {n_\p \over n_\n} \left(m-m_\p^*\right)\right]^{-1} \approx \rho_\n \ ,
\label{rhorel1}\ee
\be
\rho_{\p\p} = \rho_\p \left[ m - {n_\p \over n_\n} \left(m-m_\p^*\right) \right] \left[ m_\p^* - {n_\p \over n_\n} \left(m-m_\p^*\right)\right]^{-1} \approx \rho_\p \left( {m \over m_\p^*} \right) \ ,
\ee
and
\be
\rho_{\n\p} = - \rho_\p (m- m_\p^*) \left[ m_\p^* - {n_\p \over n_\n} \left(m-m_\p^*\right)\right]^{-1} \approx -\rho_\p \left( {m-m_\p^* \over m_\p^*}\right) \ ,
\label{rhorel3}\ee
where the approximations are valid whenever $n_\p/n_\n \ll 1$. That is, they should be accurate for neutron star cores where the proton fraction tends to be below 10\%. We will make use of these approximations in our discussion below.

\section{The entropy entrainment}

So far, most of the literature on superfluid neutron star dynamics has ignored finite temperature effects. 
The exception is a series of paper by Gusakov and collaborators \cite{gus1,gus2,gus3} [NEW?] .Their approach builds on the orthodox view of superfluidity and
makes use of the relativistic hydrodynamics prescription of Son \cite{son}.  We want to make contact with this model in order to 
infer the key finite temperature effects. In particular, we need to be able to use the results of Gusakov \& Haensel \cite{gusha1}, who
express the finite temperature effects in terms of the mass density matrix. In order to make use of these results we first of all need to translate them to the non-relativistic limit of our framework, and then elevate the model to relativity.
This will not lead to a ``complete'' model, but it is the best that we can do given the lack of microphysics information relating to the problem we are considering.

The starting point for the discussion is a three-component model, which accounts for the superfluidity of both neutrons and protons, 
and where the thermal excitations are expressed in the form of quasi-particles (index qp below). Then the total non-relativistic fluxes $j_\x^i$ can be expressed as;
\begin{eqnarray}
j^i_\n &=&
\rho_{\rm nn} \,   V^i_{\rm n}  +
\rho_{\rm np} \,  V^i_{\rm p} + (\rho_{\rm n} - \rho_{\rm nn}-
\rho_{\rm np}) \,  v^i_{\rm qp} \,, 
\label{jn}\\
j^i_{\rm p} &=& 
\rho_{\rm pp} \, V^i_{\rm p} +
\rho_{\rm pn} \,  V^i_{\rm n}+(\rho_{\rm p} - \rho_{\rm pp}-
\rho_{\rm pn}) \, v^i_{\rm qp} \ , 
\label{jp}\end{eqnarray}
(here, and in the following, we adapt the notation in such a way that the comparison with the variational description in Section~II becomes  transparent). These expressions account for the presence of entrainment through the $\rho_{\x\y}$ coefficients in the manner discussed in the previous section. It has also been assumed that the ``normal 
velocities'' 
of protons and neutrons are identical, represented by $v^i_\mathrm{qp}$ (this is a true transport velocity, in contrast to $V_\x^i$). The quasiparticles carry the entropy in the model.

The flux expressions are quite intuitive. The density of neutrons
flowing with $V_\x^i$ is $\rho_{\n\x}$ which leaves $\rho_\n-\rho_{\n\n}-\rho_{\n\p}$ to flow with $v_\mathrm{qp}^i$ (and similarly for the protons).
The quasiparticles, representing the normal fluid, are only present at finite temperatures. In the  $T=0$ limit we have
\be
\rho_\n = \rho_{\n\n} + \rho_{\n\p} \ , 
\ee
\be
\rho_\p = \rho_{\p\p} + \rho_{\n\p} \ , 
\ee
bringing us back to the expressions discussed in the previous section. At a finite temperature these relations no longer hold, the difference being the presence of 
thermal excitations. This effect was quantified by Gusakov and Haensel \cite{gusha1}, working in the normal-fluid frame.
Basically, this means that we let $v_\mathrm{qp}^i=0$ which leads to an entrainment matrix with the same functional form as in the $T=0$ case. 
The similarity is, however, deceptive as the coefficients are now temperature dependent.  In particular, the three coefficients $\rho_{\x\y}$ are  independent, so we have a three-parameter entrainment model. 

In order to compare the variational model to this description, it is useful to first consider a general frame not tied to any of the components. Identifying the quasiparticles with the entropy carrying component (the normal fluid) we  have $v_\mathrm{qp}^i=v_\s^i$. Assuming that all relative velocities are small enough that we can linearise, the momenta will be given by
\be
p_\n^i =n_\n \tilde K^{\n\n} v_\n^i + n_\p \tilde K^{\n\p} v_\p^i + s \Acal^{\s\n} v_\s^i \ , 
\label{pn}\ee
and
\be
p_\p^i =  n_\n \tilde K^{\n\p} v_\n^i + n_\p \tilde K^{\p\p} v_\p^i  + s \Acal^{\s\p} v_\s^i \ . 
\label{pp}\ee
Here we have introduced the entropy entrainment \cite{heat,cesar1},  encoded in $\Acal^{\s\x}$. We have assumed that this quantity is trivially related to the corresponding relativistic parameter (this makes sense since $\Acal^{\n\p} = K^{\n\p}$ in the cold model).  In general, the coefficients of the mobility matrix will be affected by temperature, so it will \underline{not} be the case that $\tilde{K}^{\x\y} = K^{\x\y}$. Of course, this depends on the physics involved. 
One may, for example, assume that the standard entrainment between neutrons and protons will be essentially the same as in the cold model because it originates from the strong interaction, which is only weakly temperature dependent. The coupling to the quasiparticles (represented by the entropy carrying component above) complicates the situation. We need an argument that links the three parameters $\tilde{K}^{\n\p}$, $\Acal^{\s\n}$ and $\Acal^{\s\p}$ to the $\rho_{\x\y}$ elements of the mass density matrix.

If we want to compare \eqref{pn} and {\eqref{pp} to the previous relations, \eqref{jn} and \eqref{jp}, we need to invert them to get the respective fluxes. Thus, we arrive at
\be
j_\n^i = m n_\n v_\n^i = {m \tilde K^{\p\p} \over \dettK} p_\n^i - {m \tilde K^{\n\p} \over\dettK} p_\p^i + {m\over \dettK} \left( \Acal^{\s\p} \tilde K^{\n\p} - \tilde K^{\p\p} \Acal^{\s\n}\right) s v_\s^i \ , 
\ee
where
\be
\dettK = \tilde K^{\n\n} \tilde K^{\p\p} - \left( \tilde K^{\n\p}\right)^2 \ . 
\ee
We also have
\be
j_\p^i = m n_\p v_\p^i = {m \tilde K^{\n\n} \over \dettK}p_\p^i - {m\tilde K^{\n\p} \over \dettK} p_\n^i + {m\over \dettK} \left( \Acal^{\s\n} \tilde K^{\n\p} - \tilde K^{\n\n} \Acal^{\s\p}\right) s v_\s^i \ . 
\ee
If we introduce the superfluid ``velocities'' then these relations take the form;
\be
j_\n^i = {m^2 \tilde K^{\p\p} \over \dettK}V_\n^i - {m^2 \tilde K^{\n\p} \over\dettK}V_\p^i + {m\over \dettK} \left( \Acal^{\s\p} \tilde K^{\n\p} - \tilde K^{\p\p} \Acal^{\s\n}\right) s v_\s^i \ , 
\ee
and
\be
j_\p^i = {m^2 \tilde K^{\n\n} \over \dettK} V_\p^i - {m^2 \tilde K^{\n\p} \over \dettK} V_\n^i + {m\over \dettK} \left( \Acal^{\s\n} \tilde K^{\n\p} - \tilde K^{\n\n} \Acal^{\s\p}\right) s v_\s^i \ . 
\ee
At this point, the correspondence with the fluxes in the orthodox formulation is obvious. In particular, we see that the quasiparticle contribution vanishes 
as $s\to 0$, as expected. It is also straightforward to establish that \eqref{Krel} still holds, although now for the $\tilde K^{\x\y}$ coefficients. By comparing to \eqref{jn} and \eqref{jp}, and solving for the two entropy entrainment coefficients we find 
\be 
\Acal^{\s\n} =  {m \over s \det\,\rho}\left( \det\,\rho - \rho_{\p\p}\rho_\n + \rho_{\n\p}\rho_\p\right) \ , 
\ee
and
\be 
\Acal^{\s\p} =  {m \over s \det\,\rho}\left( \det\,\rho - \rho_{\n\n}\rho_\p + \rho_{\n\p}\rho_\n\right) \ . 
\ee
This essentially completes the translation between the two models. In particular, we see how the entropy entrainment coefficients are related to the mass density matrix elements from Gusakov and Haensel \cite{gusha1}. That is, we 
have all information required for a variational finite temperature model for superfluid neutron stars. 

In order to use these results to create a workable relativistic model, it first of all makes sense to 
assume that $\Acal^{\s\x}$ is not affected by relativistic effects (which seems reasonable given the discussion in the previous section).
Next, we extend the analysis from Section~III to include the entropy component. This leads to the chemical potential being given by
\be
\mu_\n = n_\n \Bcal^\n + n_\p \Acal^{\n\p} + s \Acal^{\s\n} = m_\n c^2 + \mu_\n^\mathrm{N}\ , 
\ee
and (to leading order in the $c\to \infty$ limit)
\be
n_\n \Bcal^\n  = n_\n \tilde{K}^{\n\n} \ . 
\ee
As in the cold case this identification (together with the analogous result for the protons) allows us to infer the parameters required in the relativistic case. 
 The model is not unique, but it has the correct Newtonian limit by construction.

\section{The finite temperature entrainment matrix} 

We have shown how we can infer the entropy entrainment needed to complete the variational model from the temperature dependent
mass density matrix. This means that we can draw on the results of  Gusakov and Haensel \cite{gusha1}.
They show that the (non-relativistic) matrix elements take the form;
\be
\rho_{\n\n}(n,T) = \left[ 1- f_\n(T) \right] \bar{\rho}_{\n\n} \ , 
\ee
\be
\rho_{\p\p}(n,T) = \left[ 1- f_\p(T) \right] \bar{\rho}_{\p\p} \ ,
\ee
and
\be
\rho_{\n\p}(n,T) =\left[ 1- f_\n(T) \right]\left[ 1- f_\p(T) \right]\bar{\rho}_{\n\p} \ , 
\ee
where $\bar{\rho}_{\x\y}$ is only weakly temperature dependent~\footnote{Given the many other uncertainties in the model we may as well replace $\tilde\rho_{\x\y}$ with  the cold result $\rho_{\x\y}(n)$ from  \eqref{rhorel1}--\eqref{rhorel3}. }. The main temperature behaviour is encoded in the 
functions $f_\x$, which can be taken from \cite{yakov}. These functions vanish in the zero temperature limit and approach unity as one approaches the critical temperature for superfluidity, when $T\to T_{c\x}$. This obviously 
means that $\rho_{\x\n}\to 0$ as $T\to T_{c\n}$, reflecting the fact that all neutrons are ``normal'' in this limit (similarly for the protons).

Given this, we may consider the behaviour of the coefficients in the variational model as one of the critical temperatures is approached. 
For this exercise, let us assume that $T_{c\n}< T_{c\p}$ (this would be true near the peak of the triple gap in the example illustrated in Figure~\ref{schemes}). Introducing the shorthand notation
\be
R = \bar{\rho}_{\n\n} \bar{\rho}_{\p\p} - (1-f_\n)(1-f_\p) \bar{\rho}_{\n\p}^2 \ , 
\ee
we have
\be
\tilde K^{\n\n} = {m^2 \bar{\rho}_{\p\p} \over (1-f_\n) R} \ ,
\ee
\be
\tilde K^{\p\p} = {m^2 \bar{\rho}_{\n\n} \over (1-f_\p) R} \ ,
\ee
and
\be
\tilde K^{\n\p} = - {m^2 \bar{\rho}_{\n\p} \over R} \ ,
\ee
showing that $\tilde K^{\x\p}$ is regular as $T\to T_{c\n}$ but $\tilde K^{\n\n}$ (and hence $\Bcal^\n$ in the relativistic model) diverges.
This behaviour is in sharp contrast with the model of Gusakov et al \cite{gus1,gusha1}, where all parameters remain finite.

However, the divergence of $\tilde K^{\n\n}$ has a natural explanation in terms of the entrainment. As the system approaches the  critical temperature, the dynamics will become dominated by the thermal excitations (an increasing fraction of the neutrons are ``normal'').
Expressed in terms of the entrainment, this has the effect that it is increasingly difficult to move the neutrons relative to the entropy (the normal fluid). 
Essentially, their effective \underline{thermal} mass increases. To see this, let us assume that the entropy is locked to the proton/electron component (as in Section~II). Then the two momenta take the form
\be
p_\n^i = n_\n \tilde K^{\n\n} v_\n^i + \left( n_\p \tilde K^{\n\p} + s \Acal^{\s\n} \right) v_\p^i \ , 
\label{nmomm}\ee 
and
\be
p_\p^i = n_\n \tilde K^{\n\p} v_\n^i + \left( n_\p \tilde K^{\p\p} + s \Acal^{\s\p} \right) v_\p^i  \ , 
\ee
and we see that we can define  effective masses $\tilde m_\x^*$ that include finite temperature effects;
\be
\tilde m_\n^* = n_\n \tilde K^{\n\n} \ , 
\ee
and
\be
\tilde m_\p^* =  n_\p \tilde K^{\p\p} + s \Acal^{\s\p} \ , 
\ee
The first of these clearly diverges as the critical temperature $T_{c\n}$ is approached. The second, however, remains finite. 
This becomes clear once we note that $\tilde K^{\p\p}$ is regular and 
\be
\Acal^{\s\n} = {m\over s} \left[ 1 - {\bar{\rho}_{\p\p}\rho_\n - (1-f_\n)\bar{\rho}_{\n\p}\rho_\p \over (1-f_\n)R }\right] \ , 
\ee
and
\be
\Acal^{\s\p} = {m\over s} \left[ 1 - {\bar{\rho}_{\n\n}\rho_\p - (1-f_\p)\bar{\rho}_{\n\p}\rho_\n \over (1-f_\p)R }\right] \ , 
\ee
where the former diverges as $T\to T_{c\n}$ but the latter remains regular.

As a result of this behaviour,  the neutrons lock to the entropy (the ``normal'' component). That is,  we must have 
$v_\n^i\to v_\s^i$ in order to ensure that the dynamics remain regular. This is apparent from the fact that the neutron momentum \eqref{nmomm} takes the form
\be
p_\n^i \to  { \rho_\n \over (1-f_\n) \bar{\rho}_{\n\n}} \left(v_\n^i-v_\p^i\ \right) + \left[  m- {m\rho_\p \bar{\rho}_{\n\p} \over \bar{\rho}_{\n\n}\bar{\rho}_{\p\p}} \right] v_\p^i \ , 
\ee
as $T\to T_{c\n}$.
Demanding that the momentum remains regular simply requires the neutrons to lock to the entropy+protons in the limit. This reflects the anticipated 
physics at the critical temperature, and a reduction in the number of distinct dynamical degrees of freedom.

\section{Insights into the dynamics}

In order to explore the main features of the finite temperature superfluid model that we have developed, let us consider a
simplified dynamical problem. The main aim of this exercise is to shed light on the transition as the critical temperature is approached. 

We take as our starting point the two-fluid model that results from locking the protons  (electromagnetically) to the electrons and the entropy, cf., Section~II.
Meanwhile, the neutrons are superfluid and may move relative to the other components. We consider linear deviations from a uniform equilibrium where all fluids move together (the simplest situation). We assume that the fluids are in dynamical, chemical and thermal equilibrium in this state. This means that 
all background quantities (including the temperature) are constant (in time), and we also have $\mu_\n=\mu_\p+\mu_\e$. The equilibrium is associated with a frame such that $u^a=[1,0,0,0]$. Perturbing this system, the normal fluid will move with $u^a+\delta u^a$, where $u^a \delta u_a = 0$, and the superfluid neutrons
move with relative velocity  $\delta v_\n^a$, where $u^a \delta v_a^\n = 0$. For simplicity, we will work in the local inertial frame associated with $u^a$. That is, we assume that spacetime is flat, which means that all derivatives are partial.

Let us first consider the conservation law for entropy, which leads to
\be
\partial_t \delta s + s \nabla_i \delta u^i = 0 \ , 
\label{entrop}\ee
where we have defined the co-moving time-derivative $\partial_t = u^a\nabla_a$ and we have used the fact that the perturbed velocities are spatial ($i=1-3$). Next we perturb the stress-energy tensor and work out its divergence. This leads to
\begin{multline}
\nabla_a \delta T^{ab} = (P+\rho) u^b \nabla_a \delta u^a + (P+\rho) \partial_t \delta u^b + u^b \partial_t (\delta P + \delta \rho) \\
+ \nabla^b \delta P + n_\n \mu_\n u^b \nabla_a \delta v_\n^a + n_\n \mu_\n \partial_t \delta v_\n^b = 0 \ . 
\end{multline}
Contracting with $u_b$ we arrive at 
\be
\partial_t \delta \rho + (P+\rho) \nabla_i \delta u^i + n_\n \mu_\n \nabla_i \delta v_\n^i = 0  \ . 
\label{con1}\ee
Meanwhile, we have from the projection orthogonal to $u^b$;
\be
(P+\rho)\partial_t \delta u_i + \nabla_i \delta P + n_\n \mu_\n \partial_t \delta v_\n^i = 0\ . 
\label{eul1} 
\ee
In these two equations we recognize the standard perfect fluid terms, but also see the influence of the superfluid through the 
terms involving $\delta v_\n^i$. Next consider the neutron conservation law, which leads to
\be
\partial_t \delta n_\n + n_\n \left( \nabla_i \delta u^i + \nabla_i \delta v_\n^i \right) = 0\ , 
\label{con2}\ee
while the corresponding momentum equation can be written (the component along $u^b$ vanishes identically)
\be
\mu_\n \partial_t \delta u_i + n_\n \Bcal^\n \partial_t \delta v_i^\n + \nabla_i \delta \mu_\n = 0 \ . 
\label{eul2}\ee

Based on these equations, it is easy to show that only the longitudinal (in the spatial sense) degrees of freedom are non-trivial. Hence, we take the divergence of \eqref{eul1} and \eqref{eul2} in order to obtain scalar equations. Then we can use \eqref{con1} and \eqref{con2} to remove the explicit dependence on the 
velocities, ending up with equations relating perturbed scalar variables. The relevant equations take the form;
\be
\partial^2_t \delta \rho - \nabla^2 \delta P = 0 \ , 
\ee
and
\be
\left[ {\mu_n^2 - (P+\rho)\Bcal^\n \over n_\n \mu_\n - (P+\rho) }\right] \partial^2_t \delta n_\n - \nabla^2 \delta \mu_\n - 
\left[ {\mu_n - n_\n\Bcal^\n \over n_\n \mu_\n - (P+\rho) }\right] \partial^2_t \delta \rho = 0  \ . 
\ee
The first is, obviously, the usual wave equation for sound and the second equation is similar (even though the coefficients are a bit messier).
If we consider the case where the neutrons are not interacting with the other particles (no entrainment), then we have $\mu_\n=n_\n \Bcal^\n$ and the second equation collapses to
\be
\partial_t^2 \delta n_\n -  {n_\n \over \mu_\n} \left( {d\mu_\n \over dn_\n}\right) \nabla^2 \delta n_\n = 0 \ , 
\ee
i.e., a decoupled wave equation with wave-speed \cite{livrev}
\be
c_\n^2 =  {n_\n \over \mu_\n} \left( {d\mu_\n \over dn_\n}\right) \ . 
\ee
The general relations provide 
a neat demonstration of the existence of two identifiable wave degrees of freedom in the problem, leading to the existence of unique ``superfluid'' oscillations modes in a neutron star core \cite{Epstein,Lindb,Lin,ACmnras}.

Before moving on, it is worth noting that the entropy is slaved to the other perturbations. Solving \eqref{con1} and \eqref{con2} for the divergence of 
$\delta u^i$ and using the result in \eqref{entrop} we see that (after time integration)
\be
\delta s = {s \over P+\rho - n_\n \mu_\n} \left( \delta \rho - \mu_\n \delta n_\n \right) \ .
\label{entrel}\ee

In order to progress further, we need to introduce a three-parameter equation of state. Opting to work with the density, neutron number density and entropy density as our primary variables, we have
\be
\delta P = \left( {\partial P\over \partial \rho } \right)_{n_\n,s}\delta \rho + \left( {\partial P\over \partial n_\n } \right)_{\rho,s}\delta n_\n+ \left( {\partial P\over \partial s } \right)_{\rho,n_\n}\delta s \ , 
\ee  
and
\be
\delta \mu_\n = \left( {\partial \mu_\n\over \partial \rho } \right)_{n_\n,s}\delta \rho + \left( {\partial \mu_\n\over \partial n_\n } \right)_{\rho,s}\delta n_\n+ \left( {\partial \mu_\n\over \partial s } \right)_{\rho,n_\n}\delta s \ . 
\ee
In the following we will not state the variables that are held fixed explicitly, for the sake of clarity and brevity.
 In these relations we can use
\eqref{entrel} to replace $\delta s$, effectively reducing the problem to one with two parameters. Having done this, we can use the obtained expressions in \eqref{eul1} and \eqref{eul2}, leading to two coupled wave equations for $\delta \rho$ and $\delta n_\n$. 

These equations are rather messy, with the first  taking the form
\be
\partial_t^2 \delta \rho - a \nabla^2 \delta \rho - b \nabla^2 \delta n_\n = 0 \ , 
\label{rhoeq}\ee
where
\be
a = \left( {\partial P \over \partial \rho} \right) +  {s \over P+\rho - n_\n \mu_\n} \left( {\partial P\over \partial s } \right) \ , 
\ee
and 
\be
b =  \left( {\partial P\over \partial n_\n } \right) -  {s \mu_\n \over P+\rho - n_\n \mu_\n} \left( {\partial P\over \partial s } \right) \ .
\ee
The second equation can be written
\be
c \partial^2_t \delta n_\n - d \nabla^2 \delta n_\n - e \nabla^2 \delta \rho = 0 \ , 
\label{neq}\ee
with 
\be
c = {\mu_n^2 - (P+\rho)\Bcal^\n \over n_\n \mu_\n - (P+\rho) } \ , 
\ee
\be
d = \left( {\partial \mu_\n\over \partial n_\n } \right) -  {s \mu_\n \over P+\rho - n_\n \mu_\n} \left( {\partial \mu_\n\over \partial s } \right)
+ \left[ {\mu_n - n_\n\Bcal^\n \over n_\n \mu_\n - (P+\rho) } \right]  b  \ , 
\ee
and
\be
e =  \left[ {\mu_n - n_\n\Bcal^\n \over n_\n \mu_\n - (P+\rho) } \right] a + \left( {\partial \mu_\n\over \partial \rho } \right)
+ {s \mu_\n \over P+\rho - n_\n \mu_\n} \left( {\partial \mu_\n\over \partial s } \right) \ . 
\ee
Given these expressions it is straightforward to work out the dispersion relation for plane waves in the system. Assuming that the 
perturbations behave as $\exp(i\omega t + i k_jx^j)$ and defining the phase-velocity $\sigma=\omega/k$ we find that
\be
\left( \sigma^2 - a\right) \left( c \sigma^2 - d\right) - be = 0 \ . 
\ee
It is obviously easy to obtain the characteristic wave speeds in the system, once we introduce an explicit equation of state. We will, however, 
not do this. Instead, let us focus on two limiting cases. Consider first the low-temperature limit, i.e. when the system  far below the critical temperature, deep into the superfluid region. Then we can take the $s\to 0$  limit to obtain the dispersion relation;
\be
\left[ \sigma^2 - \left({\partial P \over\partial \rho}\right) \right] \left[ \sigma^2 - {n_\n \over \mu_\n}  \left( {\partial \mu_\n\over \partial n_\n } \right)\right] - {n_\n   \over \mu_\n} \left( {\partial P\over \partial n_\n } \right)
 \left( {\partial \mu_\n\over \partial \rho } \right) = 0 \ . 
\ee
As a sanity check of this result, it is worth noting that in the zero temperature limit we can rewrite the expression using $n_\n$ and $n_\p$ as the independent variables. Then defining
\be
c_\x^2 = {n_\x \over\mu_\x} \left( {\partial \mu_\x \over \partial n_\x} \right)_{\n_\y} \ , 
\ee
we arrive at the (more symmetric) dispersion relation
\be
\left( \sigma^2 - c_\n^2\right) \left( \sigma^2 - c_\p^2\right) - {n_\n \over\mu_\n} \left( {\partial \mu_\n \over \partial n_\p}\right) _{n_\n} 
 {n_\p \over\mu_\p} \left( {\partial \mu_\p \over \partial n_\n}\right) _{n_\p} = 0  \ . 
\ee
This result is identical to that derived in \cite{livrev}.

The other useful limit to consider corresponds to letting the temperature approach the critical temperature for the neutrons.
We know from the discussion in the previous section that the parameters in the model are singular in this limit. In particular, we know that 
\be
\Bcal^\n \sim \tilde K^{\n\n} \to \infty \ , 
\ee
while the chemical potential $\mu_\n$ remains finite.
In order for the problem to remain regular, all terms proportional to $\Bcal^\n$ in the equations  must balance.  Thus, we find  that \eqref{neq} becomes
\be
- {P+\rho \over n_\n } \partial^2_t \delta n_n + b \nabla^2 \delta n_\n + a \nabla^2 \delta \rho = 0 \ . 
\ee
Adding this to \eqref{rhoeq} (and time integrating) we see that 
\be
\delta n_\n = { n_\n \over P+ \rho} \delta \rho  \ . 
\ee
That is, one of the dynamical degrees of freedom has been quenched. The remaining wave equation can be written
\be
\partial_t^2 \delta \rho - \left[ \left( {\partial P\over \partial \rho } \right)  + {s \over P+\rho} \left( {\partial P\over \partial s } \right) + {n_\n \over P+\rho} \left( {\partial P\over \partial n_\n } \right) \right] \nabla^2 \delta \rho = 0  \ , 
\ee
which leads to the sound speed
\be
c_s^2 =\left( {\partial P\over \partial \rho } \right)  + {s \over P+\rho} \left( {\partial P\over \partial s } \right) + 
{n_\n \over P+\rho} \left( {\partial P\over \partial n_\n } \right) \ . 
\ee
To see that this is, indeed, the expected result we need to take a few more steps. First we note that the standard single fluid analysis is based on the total baryon number density $n_\mathrm{b} = n_\n+n_\p$. Secondly, it is customary to work in terms of the entropy per baryon rather than the entropy number density. Thus we define $\bar{s} = s/n_\mathrm{b}$. We can then rewrite the thermodynamical derivatives in terms of $\rho$ and $\bar{s}$. This leads to the final result
\be
c_s^2 = \left( {\partial P\over \partial \rho } \right)_{\bar{s}} \ , 
\ee
which is the standard expression for the (adiabatic) sound speed in the single-fluid case.

This model problem demonstrates that the finite temperature model we have developed connects the two-fluid region with the normal fluid region in a natural fashion. 

\section{Brief summary and outlook}

Working in the context of the variational framework for relativistic multi-fluid systems \cite{carter,livrev}, we have developed 
a model that accounts for finite temperature effects in coupled superfluid systems. 
The model builds on recent progress on the analogous problem of heat flow \cite{cesar1,cesar2} and is based on treating the thermal excitations (e.g. phonons) in the systems as an additional entropy-carrying fluid. The approach breaks down in the zero-temperature limit, where the excitations propagate ballistically rather than collectively, but should be relevant for various systems of astrophysical interest. 
Our model is designed with the dynamics of the outer core of a mature neutron star (where superfluid neutrons are coupled to a conglomerate of protons and electrons) in mind, but the main ingredients are relevant for a range of analogous problems. 

The developed formalism allows for detailed studies of the dynamics of 
neutron star superfluids near the critical density at which superfluidity sets in for a given (local) temperature. Setting the scene for more detailed work on this problem, we  explored the key role of the entrainment between the entropy and the superfluid condensates, arguing that an increasing ``thermal effective mass'' couples the components as the critical density/temperature is approached. We demonstrated this coupling in the specific case of the sound waves in the system. This exercise brings out the expected behaviour; Far below the critical temperature the systems exhibits second sound, but this phenomenon is quenched as one approaches the edge of the superfluid region. This  result has implications for future work on the dynamics of neutron star superfluids. In particular it improves our understanding of the normal to superfluid transition, a key issue for studies of, for example, realistic neutron star seismology.

 \acknowledgments
NA acknowledges support from  STFC via grant number ST/J00135X/1. \ GLC acknowledges partial support from 
NSF via grant number PHYS-0855558.

\end{document}